\documentclass[%
reprint,
superscriptaddress,
amsmath,amssymb,
aps,
twocolumn
]{revtex4-2}

\usepackage{graphicx}
\usepackage{amssymb}
\usepackage{longtable}
\usepackage{rotating}
\usepackage{array}
\usepackage{multirow}
\usepackage{makecell}
\usepackage{booktabs}
\usepackage{xcolor}
\usepackage{ulem}
\usepackage{colortbl}
\usepackage{amsmath,esint}
\usepackage[flushleft]{threeparttable}
\usepackage{hyperref}
\usepackage{tikz}

\setcitestyle{super}
\usepackage{dcolumn}
\usepackage{bm}
\usepackage{physics}
\usepackage{hyperref}
          
\newcommand*{\citen}[1]{%
  \begingroup
    \romannumeral-`\x 
    \setcitestyle{numbers}%
    \cite{#1}%
  \endgroup   
}

\begin{document}

\preprint{APS/123-QED}

\title{
Laser Driven Bulk-to-Layered Phase Transition
}

\author{Shuang Liu}
\affiliation{Guangdong Technion – Israel Institute of Technology, Guangdong 515063, China}
\affiliation{Department of Physics, Technion – Israel Institute of Technology, 32000 Haifa, Israel}
\author{Oren Cohen}
\affiliation{Guangdong Technion – Israel Institute of Technology, Guangdong 515063, China}
\affiliation{Department of Physics, Technion – Israel Institute of Technology, 32000 Haifa, Israel}
\author{Ofer Neufeld}\email{ofern@technion.ac.il}
\affiliation{Schulich Faculty of Chemistry, Technion – Israel Institute of Technology, 32000 Haifa, Israel}
\author{Peng Chen}\email{peng.chen@gtiit.edu.cn}
\affiliation{Guangdong Technion – Israel Institute of Technology, Guangdong 515063, China}
\affiliation{Guangdong Provincial Key Laboratory of Materials and Technologies for Energy Conversion, Guangdong Technion -- Israel Institute of Technology, Guangdong 515063, China}
\affiliation{Department of Physics, Technion – Israel Institute of Technology, 32000 Haifa, Israel}

\date{\today}

\begin{abstract}
Laser-induced phase transitions offer pathways of phase transitions that are inaccessible by conventional stimuli. 
In this study, we conduct {\it ab initio} simulations to numerically demonstrate a novel laser-induced structural transformation: converting a bulk crystal into a layered van der Waals material using intense light pulses. The transition is driven by a nonlinear phononic mechanism, where selectively exciting polar and anti-polar phonon modes with polarized terahertz light breaks targeted interlayer bonds while preserving intralayer ones. 
We identify that strong anisotropy in bond-sensitivity—where interlayer bonds are significantly more susceptible to excitation than intralayer bonds—is the critical prerequisite.
Our findings 
pave the way for on-demand transformations from bulk to 2D materials, facilitate the design of advanced phase-change devices, and suggest a potential optical exfoliation method to expand the range of exfoliable 2D materials.
\end{abstract}

\maketitle

The interaction between light and matter is a cornerstone of modern condensed matter physics. It has paved the way for uncovering remarkable phenomena, including long-range energy transfer\cite{light_energy_transfer_Zhong2017}, optically driven phase transitions\cite{nasu2004}, light-induced superconductivity\cite{Mitrano2016}, Floquet topological states\cite{Rechtsman2013,light_floquet_Zhang2025}, ultrafast magnetic control\cite{Kirilyuk2010,Ilyas2024}, and exciton-polariton condensation.\cite{Kasprzak2006,Kasprzak2006} In particular, intense, ultrashort laser pulses have emerged as a powerful tool for revealing exotic transition pathways that are inaccessible with conventional stimuli. By delivering an immense electric field over a very short duration—with peak fields reaching up to 100 MV/cm in the mid-infrared range\cite{Sell2008}—lasers can drive a system across large energy barriers without causing the thermal damage that a static field of similar magnitude would inflict. This capability has been harnessed to achieve remarkable, non-thermal control over material properties, which is critical for developing next-generation information devices\cite{Kampfrath2013,Basov2017}.

One of the leading strategies for such optical control is nonlinear phononics, where intense mid-infrared pulses resonantly drive specific lattice vibrations to large amplitudes\cite{Rini2007,Frst2011,Mankowsky2016,Disa2021}. 
This selected excitation of phonons can induce transient structural deformations that profoundly alter material properties\cite{Kampfrath2013}, leading to demonstrations of optically controlled functionalities such as switching on superconductivity\cite{Mankowsky2014}, inducing insulator-to-metal transitions\cite{light_insul2metal_Rini2007}, and achieving ultrafast magnetization reversal\cite{Nova2016,light_mag_switch_CrI3_Zhang2022,light_AFM2FM_CoF2_Disa2020,light_AFMreverse_Ross2024}.

\begin{figure}[!htp]
    \centering
    \includegraphics[width=1.0\linewidth]{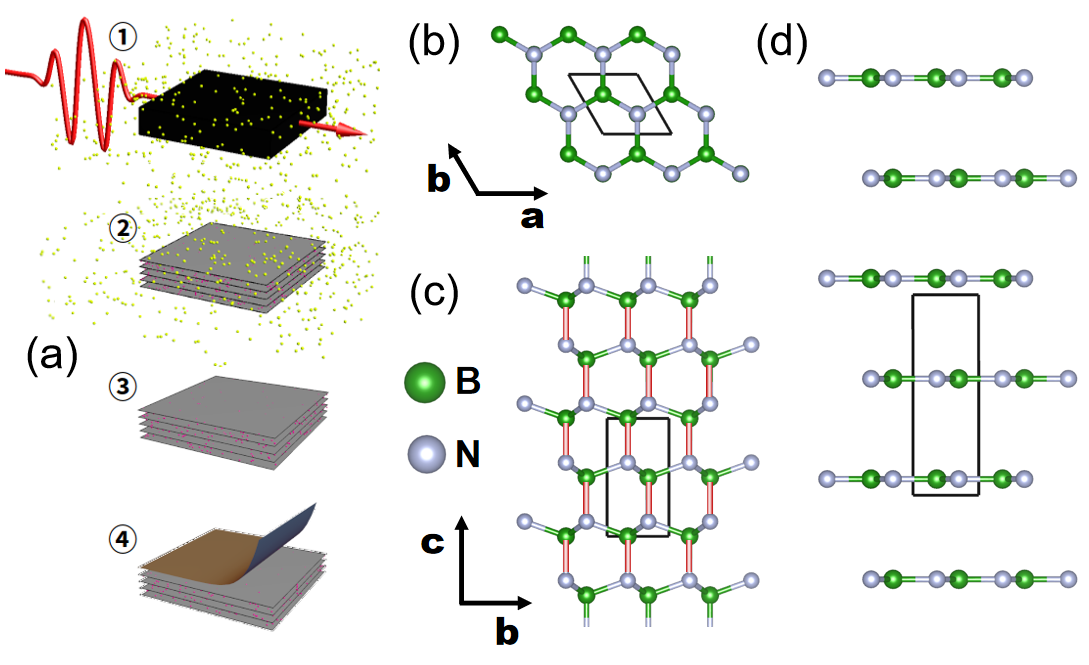}
    \caption{{\bf An illustrative diagram of the envisaged light-assisted liquid phase exfoliation and structures of boron nitride.} (a) Illustration of the envisaged light-assisted liquid phase exfoliation process: \textcircled{1} Optically excitating bulk materials within a solvent (yellow dots); \textcircled{2} Transient separation of interlayer bonds leading to the formation of thinner nanosheets; \textcircled{3} Introduction of ions (red dots) that intercalate between layers, diminishing interlayer binding and preventing re-aggregation; \textcircled{4} Retrieval of 2D nanosheets through techniques like mechanical exfoliation or centrifugation. (b) Displays the top view configurations of wurtzite boron nitride (wBN) and hexagonal boron nitride (hBN). (c) and (d) showcase a side view of wBN's bulk structure and hBN's vdw structure, highlighting out-of-plane bonds in red; the unit cell is outlined in black.}\label{fig:structure}
\end{figure}

Recently, this approach has also realised structural phase transitions, including the controlled sliding of layers in van der Waals (vdW) materials\cite{light_sliding_ferro_switch_BN&MoS2_Yang2024} and even transforming a layered vdW crystal into a bulk structure\cite{vdw2bulk_SnSe_Zhou2020}. However, the reverse process—an optically driven transition from a covalently bonded bulk structure to a layered vdW phase—remains an unexplored frontier. 
Achieving this would represent a paradigm shift, offering a possible novel technique for exfoliating 2D nanosheets directly from their bulk precursors (refer to Fig.~\ref{fig:structure} (a)). 
This approach could allow the exfoliation of materials lacking an existing van der Waals intermediate phase, thus broadening the collection of exfoliable 2D materials.

Boron nitride (BN) presents an ideal platform to investigate this possibility. As a III–V compound, its notable difference in electronegativity imparts a strong ionic character to the polar B-N bond and large Born effective charges, making it highly responsive to electric fields\cite{Geick1966,Karch1997,Wirtz2006}. BN exists in multiple polymorphs, including $\text{sp}^3$-bonded bulk forms like wurtzite (wBN) and cubic BN (cBN), and $\text{sp}^2$-bonded layered vdW structures like hexagonal BN (hBN) .\cite{BN_polymorphs_Koga2001,BN_polymorphs_Naclerio2022} While the phase transitions between these forms under conventional high-pressure and high-temperature conditions are well-understood,\cite{hBN2cBN_Biswas2024,rBN2cBN_Sato1982} the potential for optical control is just beginning to be realized, including recent experiments that have confirmed that ultrashort mid-infrared pulses can selectively break in-plane bonds in hBN,\cite{unzipping_BN_Chen2024} validating the feasibility of laser-induced bond manipulation in this material.

In this study, we demonstrate the possibility of using nonlinear phononics to selectively break the out-of-plane polar bonds in wurtzite boron nitride, thereby achieving a non-thermal, laser-driven exfoliation into its layered hexagonal phase. By {\it ab-initio} simulations, we thoroughly examine the phonon dynamics during this wBN-to-hBN transition, analyzing the energy barriers and bond evolution to elucidate the underlying physical mechanism. By identifying the optimal frequency and intensity ranges of the laser pulse required to drive this structural transformation (guiding future experiments), our work aims to establish a route for the optical control of bulk-to-layer phase transition, potentially useful for expanding the library of exfoliable 2D materials and designing phase-change memory electronics.

We perform {\it ab-initio} molecular dynamics (AIMD) simulations incorporating electromechanical responses of solids under mid-infrared laser pulses\cite{Fu2003,Wang2024,Cheng2024} to study phononic excitation and phase transitions in boron nitride as it transitions from its bulk wurtzite structure, wBN (see Figs.~\ref{fig:structure} (b) and (c)), to the van der Waals bonded layered structure, hBN (refer to Figs.~\ref{fig:structure} (b) and (d), detail see Fig. S1.), when subjected to terahertz laser pulses. 
The details of the electromechanical response theory and simulation parameters are delegated to the Supplementary Information (SI).

As illustrated in Figs.~\ref{fig:structure} (c) and (d), our simulations show that it is possible to excite a specific phonon frequency to effectively break the out-of-plane bonds (red bonds) in wBN, facilitating its conversion into van der Waals layered hBN (Details of the step-by-step structural evolution refer to the animation in the Supporting Information). 

The primary distinction between wBN and hBN lies in their bonding nature: wBN is characterized by robust sp$^3$ hybridization both in-plane and out-of-plane, whereas hBN displays sp$^2$ bonds within the plane and exhibits weak van der Waals interactions outside the plane.
This bonding characteristic leads to a significantly larger out-of-plane lattice constant $\mathbf{c}$, measuring 6.28 \AA, 
compared to 4.22 \AA, which signifies a $49\%$ tensile strain.
In addition, as illustrated in Fig.~\ref{fig:structure} (b), both wBN and hBN display a hexagonal honeycomb arrangement, yet wBN is significantly buckled, unlike the flat layers of honeycomb of hBN.

\begin{figure}[!htp]
    \centering
    \includegraphics[width=1.0\linewidth]{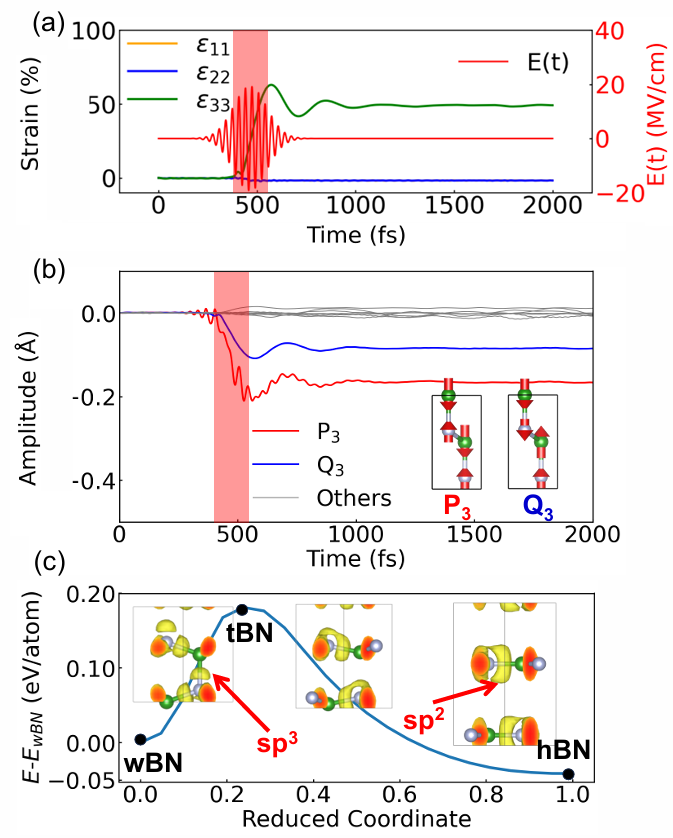}
    \caption{{\bf Phononic Excitations with Laser Pulses.} (a) the strain responses denoted as $\varepsilon_{11}$, $\varepsilon_{22}$, and $\varepsilon_{33}$, represented by the blue, orange, and green curves, respectively, along with the laser pulse's electric field over time, shown by the red curve;  (b) Temporal projections of atomic evolution on all phonon vectors; Inset illustrations highlight two primary responses: the polar and anti-polar phonon vectors, with evolution trajectory shown in red and blue, respectively, while all other trajectories are indicated in gray; (c) Energy barrier between wBN and hBN as determined from ssNEB calculations; the transitional phase with the maximal energy is denoted as tBN. The ELF, illustrating electron hybridizations, is depicted along the transition path.}\label{fig:phonon}
\end{figure}

Specifically, as illustrated by the red curve in Fig.~\ref{fig:phonon} (a), we mimic the application of a Gaussian-enveloped laser pulse of the form $E(t)=E_0\text{cos}(2\pi\omega t)~e^{-2ln(2)(\frac{t}{\tau})^2}$ (with a full-width-half-maximum (FWHM) $\tau$ = 300 fs and frequency $\omega$ = 30 THz) and a light polarization field ($\mathbf{E}$) that is 20 MV/cm and parallel to the c-axis. Results with varying light parameters are also obtained, and their findings will be summarized at the end. The AIMD is simulated for roughly 2000 fs using wBN 6$\cross$6$\cross$2 supercell as the initial structure. 

Figure \ref{fig:phonon} (a) also shows the progression of strains $\varepsilon_{11}$, $\varepsilon_{22}$, and $\varepsilon_{33}$. Alterations in the lattice constants remain minimal up until $\sim$300 fs. As the laser power peaks around 400 fs, the c-axis undergoes expansion, reaching 60\% strain, which indicates a major out-of-plane bond-breaking event, eventually stabilizing at about 50\% strain. On the other hand, the strains along the a- and b-axes decrease slightly, which is attributed to the Poisson effect.

We now delve deeply into phonon responses in order to explain the significant changes in lattice constants in response to laser irradiation. During the laser-induced phase transition, several phonon modes are activated. To evaluate their contributions to this transition, we represent atomic motions using phonon modes and extract each phonon mode's amplitude over time via a projection scheme (see details in SI). As depicted in Fig.~\ref{fig:phonon} (b), the temporal evolution of phonon mode amplitudes indicate that two main modes, namely the polar mode ($P_3$) and anti-polar mode,($Q_3$) are predominant during the dynamics. These phonons have eigen-frequencies of 30.5 THz and 31.5 THz, respectively. Phonon eigenvectors for these modes,  $P_3$ and $Q_3$, are shown as insets in Fig.~\ref{fig:phonon} (b). The anti-polar mode aligns with the symmetry of strain $\varepsilon_{33}$, allowing direct bilinear coupling, $Q_{3}\varepsilon_{33}$. On the other hand, the polar mode also interacts with the strain $\varepsilon_{33}$, represented as $P_3^2\varepsilon_{33}$, though at a higher order. Both couplings transfer energy into strain $\varepsilon_{33}$ degrees of freedom. Crucially, the $P_{3}$ mode involves the stretching of out-of-plane bonds, initiating bond breaking and changing of the lattice structure from buckled to flat layers, as observed in hBN. As a result, the polar mode is vital for facilitating the layer's flattening, and both modes enhance the c-axis strain.

In order to investigate the phase transition barrier, we utilized the solid-state nudged elastic band method (ssNEB)\cite{neb_Henkelman2000,neb2_Henkelman2000,ssneb_Sheppard2012} for identifying the minimum energy pathway from wBN to hBN. 
As depicted in Fig.~\ref{fig:phonon} (c), the total energy of hBN is approximately 0.05 eV/atom lower than that of wBN, in agreement with established literature~\cite{energy_barrier_Zhen2024}. The transition from wBN to hBN requires overcoming an energy barrier of $\sim$0.185 eV/atom, consistent with Ref.~\citen{energy_barrier_Zhen2024}. This energy barrier reveals that the phase transition should not occur spontaneously, but could be achieved under intense laser fields if those pump sufficient energy into the system in the correct pathway.

To analyze bond modifications we utilized the electron localization function (ELF)\cite{ELF_Becke1990} and examine the progression of bond structures, as illustrated in the three insets within Fig.~\ref{fig:phonon} (c). Within wBN, all bonds are sp$^3$ hybridized, demonstrating comparable geometries for both in-plane and out-of-plane bonds. In the case of tBN, the in-plane bonds exhibit increased electron sharing, indicating enhanced bonding strength, while electron sharing across the out-of-plane bonds diminishes significantly, highlighting their weakening. In the hBN structure, the out-of-plane bonds are completely broken, whereas the in-plane bonds exhibit increased electron sharing, aligning with sp$^2$ hybridized bonds.

\begin{figure}[!htp]
    \centering
    \includegraphics[width=1.0\linewidth]{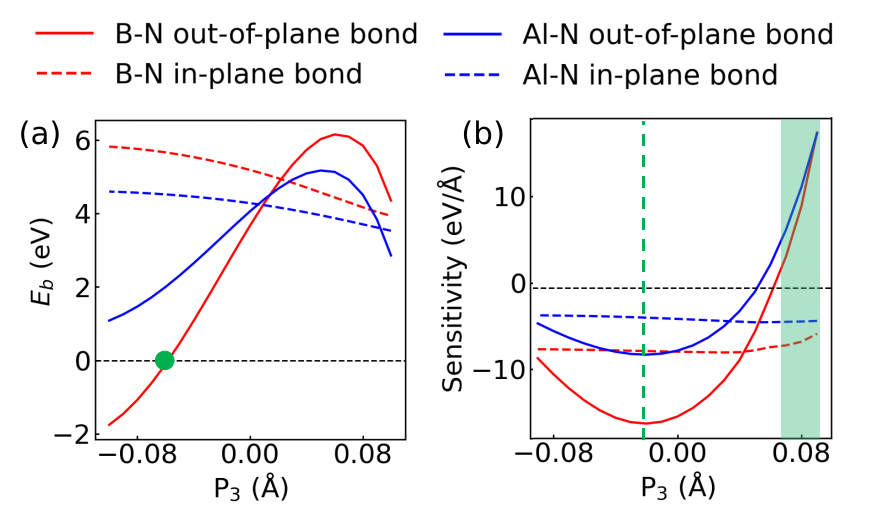}
    \caption{{\bf Sensitivity of Bond Dissociation to Phononic Vibrations.} (a) The bonding energies $E_{\text{opb}}$ and $E_{\text{ipb}}$ respect to polar mode and (b) their derivative with respect to bond length for polar modes of AlN and BN.} \label{fig:bond}
\end{figure}

Interestingly, we examined a similar effect in AlN, which has a similar wurtzite crystal structure. However, we could not optically induce a similar phase transition in our simulations (see Fig. S2 in SM). 
This is an essential data point - understanding what makes BN different than AlN regarding the phonon excitations and structural phase transitions is key to resolving the physical mechanism of the light-induced phase transition.
Thus, we investigate the impacts of the polar mode and strain individually influencing bond modifications in wBN and wurzite AlN (wAlN). We use the term ``bonding energy'' to represent the energy associated with these bonds (See detail in section IV, SI). This term will be employed to compute two types of bonding energy: one for out-of-plane bonds ($E_{\text{opb}}$) and another for in-plane bonds ($E_{\text{ipb}}$). 

With ``bonding energy'' defined, we introduce strain and polar modes to the pristine wBN and wAlN reference structure. 
We tracked the changes in bond lengths as well as in ``bonding energy'' $E_{\text{b}}$. 
As demonstrated by the solid red curve in Fig.~\ref{fig:bond}(a), the $E_{\text{opb}}$ for the out-of-plane bond decreases rapidly with an increase in negative polar mode amplitude, turning negative beyond $-0.05$ \AA~as marked by the green point in Fig.~\ref{fig:bond}(b). This indicates a bond instability and a tendency towards bond dissociation driven by the polar mode. However, the bond energy of AlN (blue solid and dashed lines in Fig.~\ref{fig:bond}(b)) remains positive throughout. Notably, the left endpoint of the blue line in Fig.~\ref{fig:bond}(b) corresponds to the bulk structure with a flat layer configuration, indicating that exfoliated layers in AlN are energetically unfavorable. In fact, even if AlN is optically deformed toward a vdW structure, it relaxes back into wAlN after the laser pulses (see more detail in Fig.~S2 (b) of the SI). Thus, optical exfoliation in BN is much favorable compared to AlN.

For a comparison and more profound understanding of the relationship between bond and strain/polar mode, we evaluate the derivatives of $E_{opb}$ and $E_{ipb}$ with respect to bond length. These derivatives, namely $dE_{opb}/dl_{bond}$ and $dE_{ipb}/dl_{bond}$, act as indicators of bond sensitivity to strain or polar mode (see more detail in section V, SI).
We directly show in Fig.~S5 of the SI that the in-plane and out-of-plane bond lengths  and opposite trends with the change of polar mode. So that the sensitivity of in-plane and out-of-plane bond correspond to polar mode has basically same sign as shown in Fig.~\ref{fig:bond} (b) (see solid and dashed lines).

As illustrated in Figs.~\ref{fig:bond} (b), the sensitivity of in-plane bonds in BN are much stronger than in AlN see dashed red and blue lines in Fig.~\ref{fig:bond}(b), both of them barely change with polar mode.
The sensitivity of out-of-plane bonds varies sharply with respect to the polar mode $P_{3}$. Specifically, as illustrated by the solid curves in the green zone of Fig.~\ref{fig:bond}(b), when the polar mode exceeds 0.04 \AA, both B-N bonds and Al-N bonds demonstrate sensitivity to changes in polarization, as evident by the steep rise in the solid curves.
However, as the polar mode decreases, B-N bonds demonstrate more pronounced changes in sensitivity compared to Al-N bonds, see solid red and blue line in Fig ~\ref{fig:bond}(b), for the polar mode near $-0.04$ \AA, the B-N out-of-plane bond sensitivity (-18.6 eV/\AA) is three times greater than that of Al-N out-of-plane bonds (-7.3 eV/\AA), as illustrated by the green dashed line in Fig ~\ref{fig:bond}(b), Such a big sensitivity corresponds to the rapidly change of BN in-plane bond energy (see red solid line in Fig ~\ref{fig:bond}(b)), and then make optically exfoliate in BN possible.

Thus, the analysis of bond energy identifies three key observations: (i) B–N bonds exhibit greater softness compared to Al–N bonds in relation to both strain (See detail in section VI, SI) and polar modes, which makes phase transitions in BN under light excitation more feasible; (ii) out-of-plane bonds react more responsively than in-plane bonds when subjected polar modes, allowing for targeted bond disruption; and (iii) polar distortions are more effective than mechanical strain at destabilizing bonds, particularly in the out-of-plane direction, highlighting the crucial influence of polar modes in light-induced phase transitions (See detail in section VI, SI).
These observations suggest that anisotropy in bond sensitivity is crucial for comprehending and forecasting optically triggered structural transformations. Additionally, they imply that the electronic bonding nature of materials can be dynamically redefined and controlled using light.

\begin{figure}[!htp]
    \centering
    \includegraphics[width=0.9\linewidth]{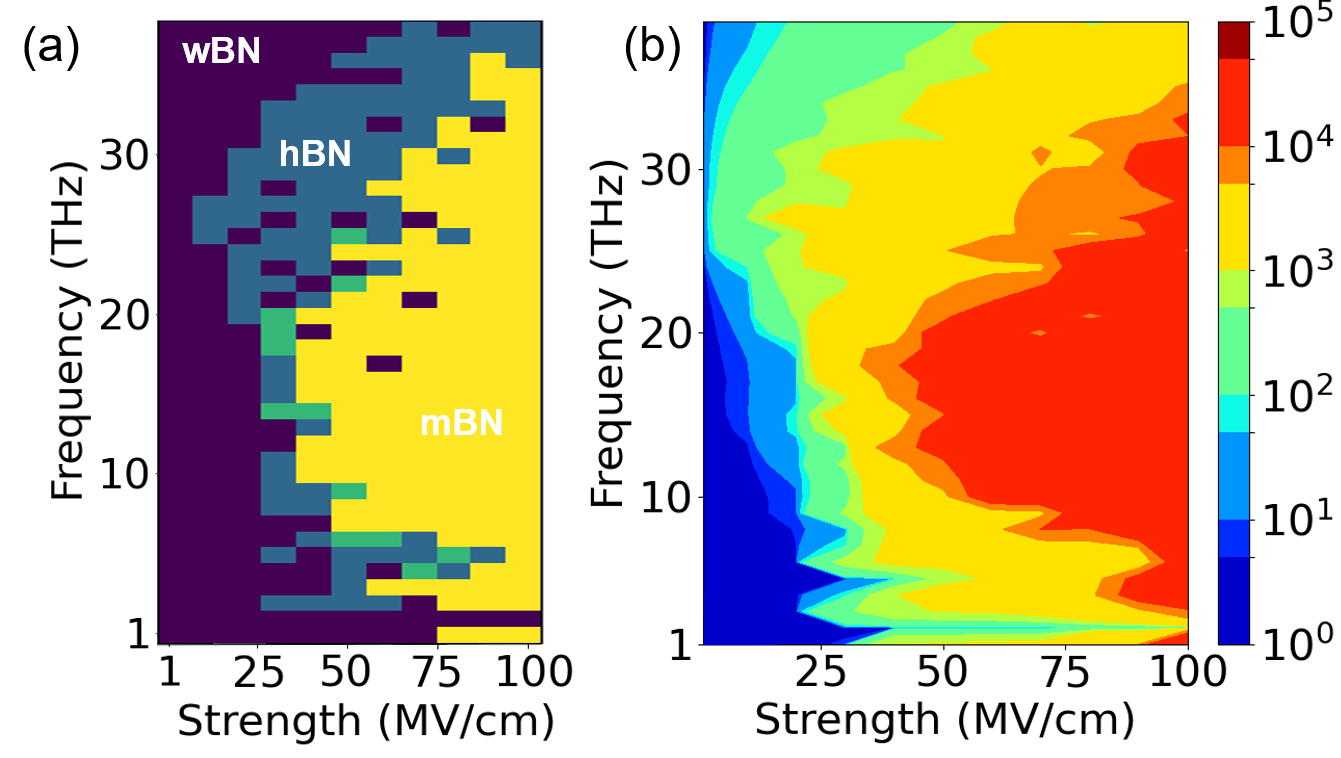}
    \caption{{\bf Phase Diagram Under Laser Pulse Parameters.} (a) different outcome with different frequencies and strengths, purple for wBN, dark blue for hBN, green for sliding hBN (sBN) and yellow for melted BN (mBN) (b) Average temperature during laser zone. }\label{fig:diagram}
\end{figure}

Lastly, to guide experiments for the optimal laser frequency and intensity for inducing phase transitions in BN, we conducted simulations scanning a frequency range of 1–40 THz and peak field strengths from 1 to 100 MV/cm. All laser pulses had an FWHM time of 300 fs. As shown in Fig.~\ref{fig:diagram}(a): wBN indicates no phase transition, colored by purple; hBN and sliding hBN (sBN) denote transitions from wBN to these respective phases, colored by deep blue and green; melted BN (mBN) refers to the complete destruction of the crystal lattice in laser zone and then reform to hBN or sliding hBN, colored by yellow.

Lasers with an electric field strength under 10 MV/cm are insufficient to trigger any phase transitions, whereas those above 50 MV/cm often result in BN melting. A significant likelihood of a phase transition manifests near 30 THz and 40 MV/cm, aligning with the frequencies of both polar and anti-polar modes, thereby validating their importance in the transition mechanism. The minimum energy threshold required for a phase transition occurs at 10 MV/cm in the 25–27 THz frequency range. Importantly, laser frequencies ranging from 10–20 THz tend to induce melting more easily.
It is important to note that our simulations are conducted under the adiabatic approximation, focusing only on phononic excitations. Electronic excitations are neglected. Consequently, for extremely high frequencies, such as those exceeding 100 THz, or in conditions of very high electric field strengths, like those beyond 50 MV/cm, our numerical simulations may not entirely capture the complexities of light-induced phase transitions. 

Figure~\ref{fig:diagram} (b) illustrates the transient temperature under various laser parameters, computed using a unit cell. Due to the constraints imposed by the small supercell in the AIMD simulation, the temperatures obtained may not be directly comparable to experimental values. Nonetheless, examining the temperature variations can provide additional insight into the phenomena occurring during the phase transition. These thermal profiles support the main conclusions derived from the phase transition map (panel (a) of fig.~\ref{fig:diagram}): (i) The primary phase transition area is concentrated around 30 THz and 40 MV/cm (dark blue region); (ii) the minimum field required for the transition from wBN to hBN is approximately 10 MV/cm and occurs at about 25–27 THz; and (iii) lasers in the 10–20 THz frequency range can easily damage the materials. By integrating the phase transition (panel (a) of fig.~\ref{fig:diagram}) and temperature (panel (b) of fig.~\ref{fig:diagram}) maps, we can conclude that shifts from wBN to hBN are likely to occur using 30 THz and 40 MV/cm laser pulses and laser peak intensities in the range 2.1 TW/cm$^2$. 

From a practical standpoint, we anticipate that our work will play an important role in advancing the understanding of BN properties under tunable lattice configurations. 
Our computational results suggest that the optimal conditions for this process include employing a laser pulse with a frequency of $\sim$30 THz with a peak electric field strength range from 10 to 50 MV/cm.
The new paradigm for controlling dimensionality with light could apply to a wide range of materials: such as phosphorus, graphite, NbN, SnSe, GeTe, and PbTe, e.g. 
Our findings provide a viable pathway for exfoliating 2D materials lacking stable vdW structures or where vdW configurations are difficult to achieve, while also give the potential for designing novel phase-change devices for next-generation electronics and in-memory computing.

In conclusion, we have theoretically demonstrated a novel class of phase transitions triggered by light, where a terahertz laser pulse changes bulk boron nitride into a layered structure by selectively exciting both polar and anti-polar phonon modes. Our ab initio simulations illustrate the crucial concept of anisotropic bond sensitivity—bonds aligned perpendicular to the planes are more reactive to polar modes, enabling selective breaking. Meanwhile, the lower sensitivity of the Al-N bond inhibits the phase transition. This insight enhances our understanding of how light-induced phase transitions occur from bulk to van der Waals structures and influence material dimensionality, which lays the groundwork for modulating these transitions using light-triggered phonons.
\\

\noindent {\bf Acknowledgment}\\
The computations were performed using the ZEUS High Performance Computing cluster at the Technion – Israel Institute of Technology.\\


\providecommand{\noopsort}[1]{}\providecommand{\singleletter}[1]{#1}%

\end{document}